\documentclass[aps,prl,reprint,superscriptaddress]{revtex4-2}
\usepackage{lipsum,amsmath,graphicx,comment,braket,xcolor,amssymb,amsfonts}
\usepackage{tikz}
\usepackage[super]{nth}
\usepackage{verbatim}

\begin{document}

\title{Field-induced jammed polyhex spin liquid in the honeycomb Ising antiferromagnet}

\author{N. Franklin}
\affiliation{Department of Physics and Astronomy, University of Manchester, Oxford Road, Manchester M13 9PL, United Kingdom}

\author{J. Richards}
\affiliation{Department of Physics and Astronomy, University of Manchester, Oxford Road, Manchester M13 9PL, United Kingdom}

\author{H. Lane}
\email[]{harry.lane@manchester.ac.uk}
\affiliation{Department of Physics and Astronomy, University of Manchester, Oxford Road, Manchester M13 9PL, United Kingdom}
\affiliation{University of Manchester at Harwell, Harwell Campus, Didcot, Oxfordshire OX11 0FA, United Kingdom}

\date{\today}

\begin{abstract}
We analyze the ground state properties of the honeycomb Ising antiferromagnet in an external magnetic field. We demonstrate the existence of extensive ground state degeneracy at finite field that maps to a polyhex tiling problem. This state is shown to be a jammed spin liquid, with no local zero modes connecting ground states. Through Monte Carlo simulations, we explore the properties of these states and show that the spin diffusion can be controlled by the magnetic field strength. By considering quantum fluctuations, we demonstrate that transverse coupling partially lifts the ground state degeneracy, selecting an extensive subspace of non-periodic tilings of 12-hexes. We suggest that the jammed polyhex spin liquid phase exists in an experimentally realizable region of parameter space and may be present in the FePX$_3$ compounds. 
\end{abstract}
\maketitle

\paragraph{Introduction --}

Studies of continuous spin systems have demonstrated many models with large classical degeneracies, including those with extensive degeneracies associated with local constraints, termed ``classical spin liquids" (CSLs)~\cite{Yan24:109,Yan24:110,lozanogomez25:preprint} and systems with subextensive degeneracies known as``spiral spin liquids"~\cite{mulder10:81,Reuther14:90,Niggemann20:32,Yan22:4}. In contrast, discrete Ising spins preclude the formation of spiral states, which typically minimize the energy in continuous spin models with competing interactions. Instead, short-range or complicated spin configurations can arise~\cite{Villain81:42,Moessner00:84,Selke88:170,Fisher80:44,Wannier50:79,Jin12:108,Bramwell01:294}.

When combined with disorder, competing interactions can promote metastable states and slow dynamics~\cite{Binder86:58}. This glassy behavior is typically driven by quenched disorder which creates a rugged energy landscape, however recently a ``spin jam" has been proposed in a kagom{\'e} model with a large classical degeneracy but whose quantum fluctuations induce metastability~\cite{Klich14:5}. These states resemble ``jammed spin liquids"~\cite{Bilitewski17:119,Bilitewski19:99} (named in analogy to jammed granular media~\cite{Parisi10:82}), whose parent Hamiltonians hosts spin liquids, but whose energy landscape is disrupted by quenched disorder.

In some frustrated models, the simplest degrees of freedom are no longer the spins, but color charges~\cite{Yan25:7,Kondev95:52,lozanogomez24:15}, dimers~\cite{Rokhsar88:61,Moessner01:63,Sachdev89:40,Moessner01:64} or plaquettes~\cite{Leung96:54,Moessner01:64,Ivanov04:70,Senthil00:62}. The tools of combinatorics can be used to characterize the degeneracy, by counting all possible colorings or tilings of the lattice. In this paper, we demonstrate a connection between polyomino tiling problems (of the kind popularized in Scientific American's ``Mathematical Games'' column~\cite{Gardner67}) and a new type of spin liquid state in the honeycomb Ising model. 

We suggest the presence of a new type of spin liquid stemming from a realistic microscopic Hamiltonian, namely the $J_1$-$J_2$-$J_3$ Ising model in field. This state comprises disordered arrangements of ferromagnetically-aligned dimers $\uparrow \uparrow$ against a fully polarized $\downarrow$ background. In contrast to the jammed spin liquid in Refs.~\cite{Bilitewski17:119,Bilitewski19:99}, this state appears in the absence of quenched disorder. This state is also distinct from the ``spin jam"~\cite{Klich14:5,Yang15:37,Samarakoon16:113,Piyakulworawat24:109} since its degeneracy is extensive and exists at the level of the classical theory. 

We show that ground states can be represented as tilings of $n=4$ polyhexes, known as ``bee polyhexes"~\cite{Gardner67}. The polyhexes admit an infinite non-periodic tiling of the plane and hence an extensive degeneracy, but the model possesses no zero modes, leading to jamming. We demonstrate that close to the transition, the field plays the role of a chemical potential, with faster dynamics at low bee populations as the jamming becomes less pronounced. Finally, we show that quantum fluctuations select an extensive subspace leading to a large ground state superposition of non-periodic tilings of 12-hexes. Our results suggest that frustrated Ising systems may provide an interesting playground, where large ground state degeneracies can be formed by polyomino tilings.

\paragraph{Model --}
We begin by introducing the Ising model in applied field
\begin{equation}
    \mathcal{\hat{H}} = \sum_{\langle i,j\rangle_n }J_n S_i^zS_j^z +h\sum_i S_i^z.
\end{equation}
\noindent Angular brackets correspond to the $n^\text{th}$ neighbors on the honeycomb lattice. Spin variables are taken to be of unit length $S_i^z \in \{-1,1\}$. In this paper we take the case where $J_1 <0$ represents a ferromagnetic interaction, as is typical in material realizations~\cite{Wildes12:24,Wildes20:101,Wildes23:107,Chen24:9,LeMardele24:109}, and set $J_1 = -1$ as a global energy scale. This model has been studied analytically by Kud\={o} and Katsura~\cite{Kudo76:56} and numerically by Wildes \textit{et al}~\cite{Wildes20:101}. 
\begin{figure}
    \centering
    \includegraphics[width=1\linewidth]{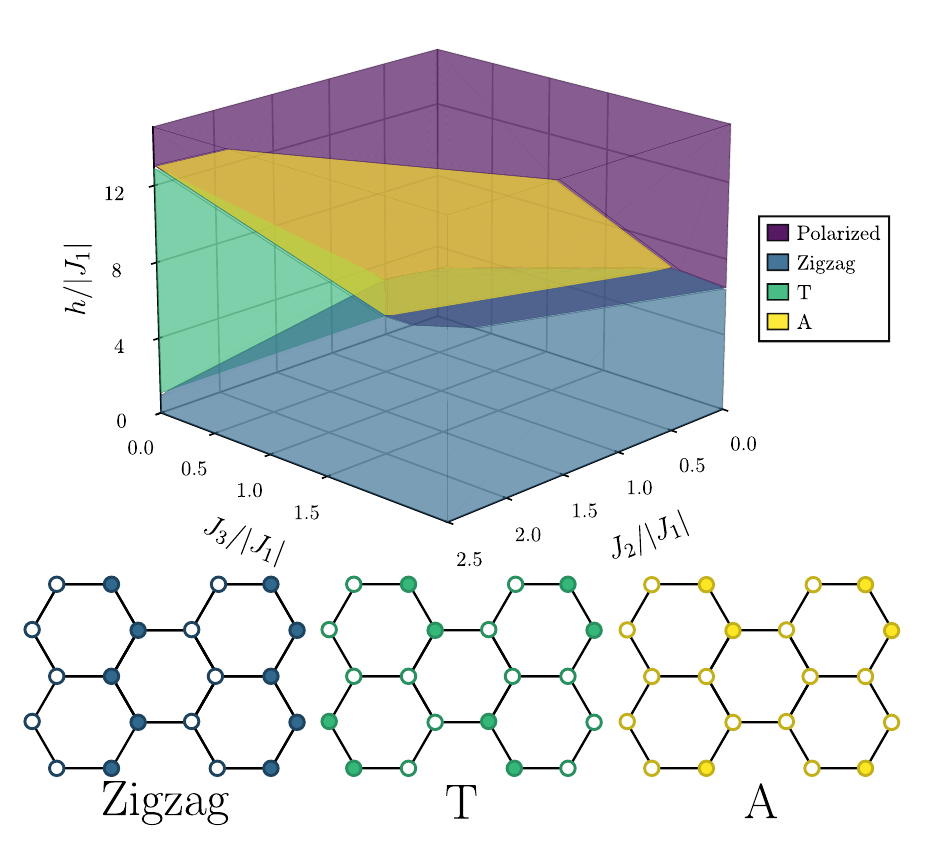}
    \caption{Phase diagram of the Ising Honeycomb model, based on the analytical results of Ref.~\cite{Kudo76:56}. Down (up) spins are represented by open (filled) circles.}
    \label{fig:fig1}
\end{figure}
In addition to the zero-field zig-zag, stripe, armchair, N{\'e}el, and ferromagnetic phases, field-induced states, consistent with $M/M_s = 1/3$ and $1/2$ plateaus are predicted (Fig.~\ref{fig:fig1}).

In this work, we focus on $J_2, J_3 >0$ (antiferromagnetic) where the classical phase diagram contains a field-induced $M/M_s=1/2$ phase, which we refer to as the A phase. The unit cell comprises six $\downarrow$ spins and two $\uparrow$ spins with an energy per site of $E_A=S^2J_1/2-hS/2$. The contributions from $J_2$ and $J_3$ cancel and the energy depends only on the nearest neighbor couplings. The A state can be viewed as a polarized $\downarrow$ background with a regular arrangement of $\uparrow\uparrow$ dimers, where all $\uparrow$ spins are connected through $J_2$ and $J_3$ to $\downarrow$ spins (Fig.~\ref{fig:fig1}). Therefore, it is natural that this state is stabilized in the quadrant associated with antiferromagnetic $J_2$ and $J_3$. Another state appearing in this quadrant is the $1/3$ plateau state, T, which corresponds to chains of $\uparrow\uparrow$ dimers connected to each other by $J_3$.

For our regime of interest, it is useful to establish the $\uparrow\uparrow$ dimer as the fundamental unit of the problem. The energy per spin of the polarized state is $E_{\mathrm{pol}}=S^2(3J_1+6J_2+3J_3)/2-hS$. Introducing a single spin flip pays the penalty $\Delta E_\uparrow=-2S^2(3J_1+6J_2+3J_3)+2hS$. For ferromagnetic $J_1$ and antiferromagnetic $J_2$, $J_3$, isolated spin flips are unstable to the formation of a dimer, $\Delta E_\text{dimer} <2\Delta E_\uparrow$.

\begin{figure}
    \centering
    \includegraphics[width=1\linewidth]{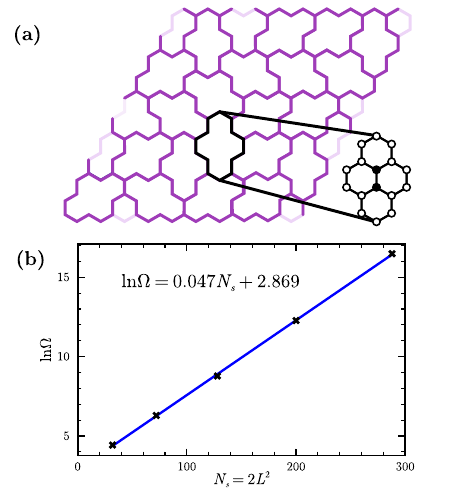}
    \caption{(a) One possible tiling on a $(10,10)$ supercell with PBCs. Edges of the supercell which are within a periodically wrapped polyhex are shaded light purple. Indicated is the spin configuration on a single polyhex, with solid (empty) circles representing up (down) spins. (b) Scaling of ground state degeneracy with system size. }
    \label{fig:fig2}
\end{figure}

\paragraph{Tiling problem --} In the A state, a convenient representation is to map onto a tiling problem. The dimers can be represented by $n=4$ symmetric polyhexes, known in recreational mathematics as ``bee polyhexes" (Fig.~\ref{fig:fig2}a). The bees are created by fusing the four hexagons surrounding the dimer such that the edges of the polyhex join all spins that are coupled to the dimer. The A state can be viewed as a regular tiling of bee polyhexes. 

Because all bonds connected to the dimer are contained within the polyhex, the energy is independent of the polyhex tiling. The A state is just one of a potentially large set of degenerate ground states represented by different tilings. The enumeration of degenerate states becomes a question of counting the ways of infinitely tiling the plane. Here, we will demonstrate that the number of tilings scales extensively and estimate the ground state degeneracy. 
To examine the degeneracy, we search for exact tilings of an $L\times L$ supercell of hexagonal tiles by bee polyhexes. This was achieved by generating a list of ways each hexagon can contribute to a polyhex and iterating over the hexagons, first selecting the hexagon with the fewest possibilities and choosing a polyhex that tiles that cell from the list of remaining polyhexes. This was repeated until an exact tiling of the supercell was found or until no available placements exist. Backtracking~\cite{Golomb65:12,Scott58:1} was implemented to explore other branches of the tiling. This procedure follows the logic of Knuth's Algorithm X~\cite{Knuth00:preprint}.

Assuming non-periodic boundaries, we find only a single tiling, corresponding to the A state (Fig.~\ref{fig:fig1}). The imposition of hard boundaries on the parallelogram is too restrictive to exhaust all tilings. With periodic boundary conditions (PBCs), we can capture tilings for which spaces are left at the edges of the parallelogram which can accommodate further tiles~\footnote{This leads to an under-count since there is no requirement that notches and protrusions on opposite side of the supercell tessellate in the limit that the supercell volume tends to infinity.  }. From this we can characterize the scaling of ground state degeneracy. Fig.~\ref{fig:fig2} shows that the logarithm of the tiling degeneracy for supercells of $N_s=2L^2$ spins.

The ground state degeneracy scales exponentially with system size $N_s=2L^2$ according to $\Omega =Ae^{\gamma N_s}$ ($A=17.628$~\footnote{Our procedure treats states that differ by the symmetry elements of the supercell as distinct. Inclusion of four symmetry elements reduces $A$ by a factor of four but does not affect the scaling.},$\gamma=0.047$), confirming the extensive ground state degeneracy~\footnote{This scaling fails in the limit of small $N_s$ where the polyhex size is comparable with the supercell size.}. This result also suggests the existence of an infinite non-periodic tiling of the plane by bee polyhexes.  

\paragraph{Thermodynamics --} The set of polyhex tilings have an extensive degeneracy, in common with a conventional CSL, yet the Ising nature of the spins precludes zero modes. This state is therefore distinct from a conventional CSL and is more alike the spin jam~\cite{Klich14:5} or jammed spin liquid~\cite{Bilitewski17:119,Bilitewski19:99} states on the kagom{\'e} lattice, although neither quantum fluctuations nor quenched disorder are required to remove the zero modes.

To further explore this polyhex spin liquid phase, we performed Monte Carlo on a $36 \times 36$ supercell with PBCs. Parallel tempering~\cite{Swendsen86:57} was implemented with 150 replicas to overcome local energy minima. Calculations were performed using the \textsc{Sunny.jl} package~\cite{Sunny:software}.
\begin{figure}
    \centering
    \includegraphics[width=1.\linewidth]{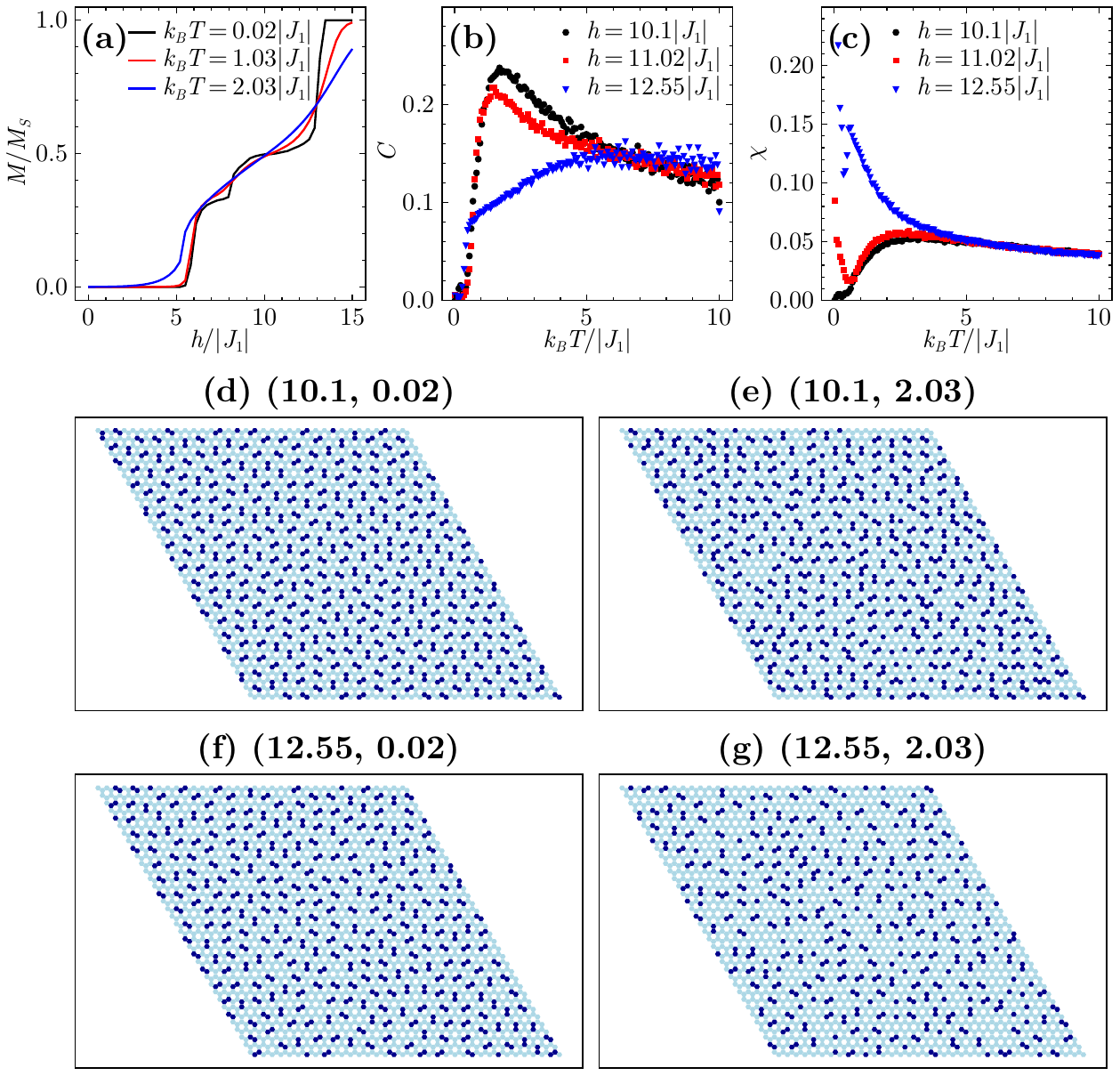}
    \caption{(a) Magnetization as a function of temperature showing the emergence of the $M/M_S = 1/2$ plateau. (b) Specific heat and (c) magnetic susceptibility showing the paramagnetic to polyhex spin liquid transition. (d-g) Representative spin snapshots at different $(h/|J_1|,k_B T/|J_1|)$ values. }
    \label{fig:fig3}
\end{figure}
We fixed $J_2 = 2J_3=2|J_1|$ for which there is a series of transitions: zigzag $\to$ T $\to$ A in field (Fig.~\ref{fig:fig3}). At low temperatures we see a transition to a polyhex spin liquid state marked by a plateau at $M/M_s = 1/2$. The transition is first order and shows clear features in magnetization, specific heat and magnetic susceptibility (Figs.~\ref{fig:fig3}a-c). 

With increased field, the transition moves to lower temperature, approaching the analytical critical field $h_c = 2J_1+6J_2+3J_3 = 13|J_1|$~\cite{Wildes20:101,Kudo76:56}. The transition from the $1/2$ plateau to the polarized state has a finite slope, at all temperatures considered. The slope can be viewed as arising due to a depletion of bee polyhexes on the lattice as the critical field is approached. Such behavior is to be expected in the vicinity of a first order transition and can be viewed as phase coexistence between a polyhex tiling and the polarized phase. This is further underscored in Fig.~\ref{fig:fig3}d-g where representative snapshots of the spin configuration are plotted at two fields on the $1/2$ plateau, showing a difference in bee population. At high fields, approaching the field-induced transition to the polarized state, the low temperature configuration is a depleted tiling (Fig.~\ref{fig:fig3}f) with voids between some polyhexes. Above the transition temperature, spin dimers are seen, along with single spin flips and local correlations that are reminiscent of both the $1/3$ and zigzag states. 

\paragraph{Dynamics --} We now consider the polyhex spin liquid dynamics. Although the ground state has an extensive degeneracy, the discrete nature of the spins precludes the existence of zero modes connecting nearby states in this space. In fact, the dynamics are \textit{severely} constrained. 
\begin{figure}
    \centering
    \includegraphics[width=1.\linewidth]{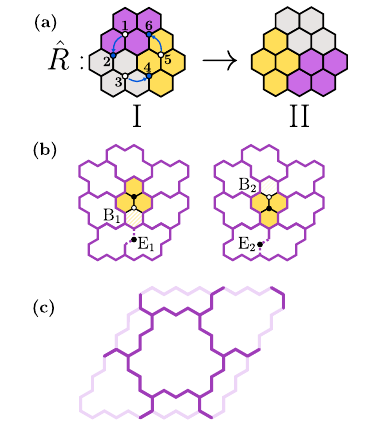}
    \caption{(a)  Two tilings of a 12-hex region, $S$, differing by the action of the operator $\hat{R}$. The I domain can be transformed to the II domain by the action of $\hat{R}$. (b) Action of a spin flip on a polyhex tiling. $\text{B}_1$ and $\text{B}_2$ correspond to two degenerate states created by acting on the bee polyhex. $\text{E}_1$ and $\text{E}_2$ are degenerate edge excitations. (c) One possible tiling of a $6 \times 6$ supercell by three 12-hexes.}
    \label{fig:domains}
\end{figure}
For a given bee polyhex tiling of the plane, $C$, consider a region of the honeycomb lattice, $S$, containing an integer number of polyhexes. If the polyhex tiling of $S$ is a subset of the infinite polyhex tiling of the plane, $A \subset C$, then any complete tiling of $S$ leads to a degenerate ground state tiling. If we consider a set of local operations in $S$ that transform between complete tilings of $S$, $ \hat{R}: A \to A'$, the action of these operations also leads to a new complete tiling of the plane, $ \hat{R}: C \to C'$. 

Since the parallelogram with hard boundaries has only a single tiling, we search for an irregular region with a tiling degeneracy. A region containing $\leq 2$ polyhexes has only a single possible tiling. In contrast, three polyhexes can be arranged in an irregular hexagon which admits two possible tilings with distinct patterns (Fig.\ref{fig:domains}a). The local operation linking these two nearby tilings requires six spin flips.

\begin{figure}
    \centering
\includegraphics[width=1.\linewidth]{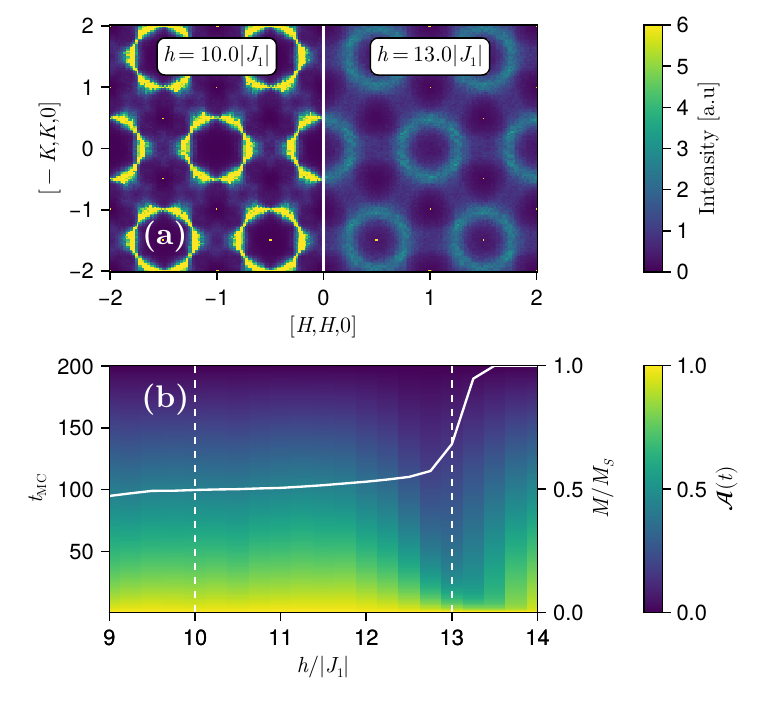}
    \caption{(a) $S(\bf{Q})$ at $k_BT=0.75|J_1|$, showing diffuse patterns that broaden near the transition. Intensities are seen at $\Gamma$ due to the finite magnetization in the A phase.  (b) Autocorrelation $\mathcal{A}(t)$ as a function of field at $k_BT=0.75|J_1|$. Dashed lines correspond to the field values in (a). Overlaid is the magnetization curve for this temperature. }
    \label{fig:dynamics}
\end{figure}

To explore the dynamics of the polyhex spin liquid phase we performed Monte Carlo simulations and calculate the static structure factor
\begin{equation}
    S(\mathbf{Q}) = \frac{1}{N}\sum_{ij}\langle S_iS_j\rangle e^{-i\mathbf{Q}\cdot \mathbf{r_{ij}}}
\end{equation}
\noindent sampling the spin configuration 100 times. The result is plotted in Fig.~\ref{fig:dynamics}a for $h=10|J_1|$ and $h=13|J_1|$. Diffuse bow-tie-like patterns are observed centered on the $M$ points. These patterns become broad rings at $h=13|J_1|$ as the bee polyhexes are depleted and become more mobile. In Figure~\ref{fig:dynamics}b the autocorrelation function 
\begin{equation}
    \mathcal{A}(t) = \frac{1}{N}\sum_i \langle S_i(t)S_i(0)\rangle
\end{equation}
is plotted in the vicinity of the $M/M_S =1/2$ plateau. Despite the extensive ground state degeneracy, the autocorrelation remains large for long times, reflecting the jamming. Towards the magnetization step at $h=13|J_1|$, the autocorrelation decays faster, signaling an increase in spin diffusion as the bee polyhexes are depleted and their motion is less constrained. Finally, in the polarized phase, the autocorrelation function recovers and the spins crystallize in the polarized arrangement.

The presence of phase coexistence at the boundary between the polyhex spin liquid and the polarized phase implies that field serves as an effective chemical potential for the creation of polyhexes from the polarized phase (polyhex vacuum), allowing some control over the rate of spin diffusion. 

\paragraph{Excitations -- }
The excited states created by a spin flip fall into two categories. The first, are states in which the total magnetization is increased by $2 S/N$ by flipping a single dimer spin. In the language of polyhexes, an $n=3$ ``stone" polyhex~\cite{Rhoads05:174} is left behind, along with an uncovered void. In contrast to the ground state space, nearest-neighbor transverse coupling is sufficient to move between these degenerate excited states. The void can be moved at zero cost by flipping a pair of spins, however this motion is highly constrained such that only the symmetric ``stone" polyhexes are created (see $\text{B}_1$ and  $\text{B}_2$ in Fig.~\ref{fig:domains}b) . The second type of excitation is the flip of a spin on the polyhex edges (see $\text{E}_1$ and  $\text{E}_2$ in Fig.~\ref{fig:domains}b). The energetics of these states depend on the local polyhex configuration. Although these excitations reduce the magnetization by $2 S/N$, for some parameterizations the energy gain from $J_2$ and $J_3$ offsets the Zeeman cost and these states are lower in energy than ``stone" excitations. For both excitation types we expect the spectrum to be localized, as energy barriers prevent the motion of spin flips freely throughout the lattice, except in the special case where the two excitations are degenerate and excitations can hop from edge to dimer.

\paragraph{Quantum fluctuations --}
We conclude with a discussion of the behavior away from the classical limit. Quantum fluctuations can be included through the transverse spin exchange terms
\begin{equation}
    \mathcal{\hat{H}}'=\frac{1}{2}\sum_{\langle ij\rangle_n}J^{\pm}_n(\hat{S}_i^+\hat{S}_j^- + h.c)
\end{equation}
\noindent which induce dynamics at zero temperature. Assuming $J^{\pm}_n$ is small, we can perform degenerate perturbation theory to derive a Hamiltonian within the low energy subspace, $C$,
\begin{equation}
    \hat{\mathcal{H}}_\text{eff} = (1-\mathcal{P})\left[\mathcal{ \hat{H}}'-\mathcal{\hat{H}}'\frac{\mathcal{P}}{\mathcal{\hat{H}}_0}\mathcal{\hat{H}}'+\mathcal{\hat{H}}'\frac{\mathcal{P}}{\mathcal{\hat{H}}_0}\mathcal{\hat{H}}'\frac{\mathcal{P}}{\mathcal{\hat{H}}_0}\mathcal{\hat{H}}'\right]
(1-\mathcal{P})\end{equation}
\noindent where the operator $\mathcal{P}$ projects onto the orthogonal complement of the ground state space, i.e. $1-\mathcal{P} = \mathcal{P}_C$ is the projector onto $C$. The first order term vanishes. At second order, couplings for all coordination shells contribute since spin flips can hop across a bond and back, returning to the ground state. At third order, $J_2^{\pm}$ gives rise to a trivial energy shift from the hopping of a spin flip around the $A(B)$ sublattices of a hexagon. At third order, $J_2$ can also facilitate a hopping which converts between domain I and domain II (see Fig.~\ref{fig:domains}a)
\begin{equation}
\begin{split}
    \mathcal{H}_\text{ring} =& J_\text{ring}\sum_R (S_1^-S_2^+S_3^-S_4^+S_5^-S_6^+ + h.c) \\=&  J_\text{ring}\sum_R  (\hat{R} +\hat{R}^\dagger )
\end{split}
\end{equation}
\noindent where $J_\text{ring}$ is comparable with the hexagonal ring exchange term in the $U(1)$ spin liquid phase of the pyrochlore antiferromagnet~\cite{Hermele04:69}. This term hybridizes states which differ by the tiling of the region $S$, whilst lowering the energy (see Fig.~\ref{fig:domains}a). Therefore, quantum fluctuations select states that comprise tilings of domains I and II. We remark that the previously predicted state A is not among the states in this subspace.   

We have demonstrated that at third order, quantum fluctuations select superpositions that comprise tilings of domain I and II, which themselves form 12-hexes. Whilst this is a subset of all infinite non-periodic tilings, it still possesses an extensive degeneracy. This is clear if one considers a parallelogram of dimensions $6 \times 6$ with PBCs (Fig.~\ref{fig:domains}c). The irregular hexagon (Fig.~\ref{fig:domains}a) has two inequivalent orientations related by $180^\circ$ rotations. The periodic $6 \times 6$ supercell can be completely tiled with three of our 12-hexes, with one 12-hex not crossing any of the boundaries (Fig.~\ref{fig:domains}c). Since there is a two-fold degeneracy in tiling a given 12-hex with three bee polyhexes, the plane can host an infinite non-periodic tiling of the domains I and II. 

\paragraph{Conclusion --} In this paper we have demonstrated the presence of a new type of spin liquid in the Ising honeycomb antiferromagnet. The ground state space is described by infinite non-periodic tilings of bee polyhexes. The dynamics are highly constrained, with ground states separated by large energy barriers, in a phenomeon similar to jamming. Finally, we demonstrated that quantum fluctuations select an extensive subset of states formed by superpositions of non-periodic tilings of the domains I and II. We stress that the parameter region in which we expect a polyhex spin liquid is large, and overlaps with the expected parameter regime of known materials, such as FePS$_3$~\cite{Wildes20:101}, where an $M/M_s$ plateau has been observed.

\begin{acknowledgments}
We are grateful to Itamar Kimchi for his insightful comments on a draft of this manuscript. Computing resources were provided by the STFC Scientific Computing Department’s SCARF cluster.
\end{acknowledgments}

\bibliography{references}

\end{document}